\documentclass[pra,twocolumn,showpacs,preprintnumbers,floatfix]{revtex4-1}
\usepackage{graphicx}% Include figure files
\usepackage{dcolumn}% Align table columns on decimal point
\usepackage{bm}% bold math
\usepackage{latexsym,epsfig}
\usepackage{graphicx}
\usepackage{verbatim}
\usepackage{comment}
\usepackage{amsmath}
\usepackage{amssymb}
\usepackage{stmaryrd}
\usepackage{color}

\newcommand{\beq}{\begin{equation}}
\newcommand{\eeq}{\end{equation}}
\newcommand{\bea}{\begin{eqnarray}}
\newcommand{\eea}{\end{eqnarray}}
\newcommand{\ben}{\begin{eqnarray*}}
\newcommand{\een}{\end{eqnarray*}}
\newcommand{\bfig}{\begin{figure}}
\newcommand{\efig}{\end{figure}}

\begin{document}
\title{Quantum phases of attractive bosons on a Bose-Hubbard ladder with three-body constraint}
\author{Manpreet Singh$^1$, Tapan Mishra$^2$, Ramesh V. Pai$^3$ and B. P. Das$^{1}$}
\affiliation{$^1$Indian Institute of Astrophysics, II Block, Koramangala, Bangalore-560 034, India}
\affiliation{$^2$Institut f\"{u}r Theoretische Physik, Leibniz Universit\"{a}t Hannover, Appelstr. 2, 30167 Hannover, Germany}
\affiliation{$^3$ Department of Physics, Goa University, Taleigao Plateau, Goa 403 206, India}

\date{\today}

\begin{abstract}
We obtain the complete quantum phase diagram of bosons on a two-leg ladder in the presence of
attractive onsite and repulsive interchain nearest neighbor interactions 
 by imposing the onsite three body constraint. We find three distinct phases, namely,
 the atomic superfluid(ASF), dimer superfluid (DSF) and the dimer rung insulator (DRI).
In the absence of the interchain nearest neighbor repulsion, the system exhibits a transition
 from the ASF to the DSF phase with increasing onsite attraction. However, the presence of the interchain
 nearest neighbor repulsion stabilizes a gapped DRI phase, which is flanked by the DSF phase. We also obtain
  the phase diagram of the system for different values of the interchain nearest neighbor interaction. By evaluating different
 order parameters, we obtain the complete phase diagram and the properties of the phase transitions
 using the self consistent cluster mean field theory.

\end{abstract}


\maketitle

\section{Introduction}
Systems of ultracold bosonic atoms in optical lattices have acquired a lot of importance from several areas of physics, in
particular in the field of condensed matter and AMO physics~\cite{lewenstein_book}. The flexibility in controlling the system
parameters by tuning the laser intensity and the technique of Feshbach resonance, makes them a versatile tool for
simulating many interesting physical systems. This results in the observation of various quantum phase
transitions which would otherwise be very difficult to study using solid-state systems.
After the first prediction ~\cite{jaksch} and observation ~\cite{bloch} of the superfluid (SF)-Mott insulator (MI) transition
in the Bose-Hubbard model in an optical lattice, enormous progress has been made in the last decade or so. It has been proposed
that a system of polar gases in optical lattices can give rise to a crystalline phase~\cite{pfaureview} due to the long
range van der Waal type interaction. Under certain conditions, it is possible to stabilize the
exotic supersolid phase~\cite{dorneich,kampf,prokofev,troyer,damle,balents,scalettar,luthra,3_10mypaper}. The observation of chromium
Bose-Einstein condensate (BEC)~\cite{Griesmaier,Beaufils15mypaper} followed by
the realization of quantum gases in other highly-magnetic
species, including dysprosium Bose and Fermi gases~\cite{16mypaper}
and erbium condensate~\cite{17mypaper}, experiments on polar molecules such as KRb~\cite{18mypaper} and
the Rydberg gases~\cite{21mypaper} have opened up possibilities for manipulating the
off-site interactions in optical lattices.

Low dimensional systems have been studied widely in the past few decades
In particular, $1D$ or quasi-$1D$ systems are of very special interest because
interactions play a crucial role in realizing novel phases~\cite{giamarchi_book,rigol_giamarchi_rmp}. Research on ultracold
atoms in optical lattices offers a unique platform to engineer various lattice models which can mimic physical
phenomena~\cite{rigol_giamarchi_rmp}. Quasi-$1D$ systems such as ladders have been of special interest
to understand the phenomenon of high-temperature
superconductivity, spin-gapped metallic state ~\cite{dagotto,uehara,kim} etc. The extra coupling between the
legs of the ladder makes these systems unique, as a result of which, the quantum phase transitions are
influenced substantially even in a simple model like the Bose-Hubbard ladder~\cite{giamarchi1,giamarchi2,meetu}.
Also, the effect of kinetic frustration along with various interactions 
can lead to interesting  new phases in ladder systems
~\cite{giamarchi_meinereffect,dhar1,dhar2,dhar3,mishra1,mishra2,lehur,altman,arun},
which are not possible in one dimensional lattice systems. In recent
years, it has been shown that novel quantum
phases such as a trimer liquid and the devil's staircase can arise from atoms and molecules possessing long range
interactions~\cite{dalmonteprl,parish_prl}.
In the experiments ladder models can be realized in the systems of optical
lattices ~\cite{danshita1,danshita2,albiezprl_2005,bloch2014}.

Earlier studies have shown that an ultracold bosonic gas in a lattice with attractive interactions undergoes a
transition from an atomic superfluid (ASF) to the dimer superfluid (DSF) phase when the atoms are subjected to the onsite three body
constraint~\cite{daley1}. This phenomenon was first predicted in the context of an atomic Bose gas in the
continuum with Feshbach resonance~\cite{stoodprl_2004}. This prediction suggests that the
bosons can pair up to form the DSF phase when
the attraction between them is sufficiently large. This transition was predicted to be Ising like at the
commensurate filling and first order at other fillings. Detailed investigations on this model have been made on a square
lattice in the recent past~\cite{wessel,lee} to obtain the ground state phase diagram. The effect of nearest neighbor
interaction on a square lattice has predicted a dimer checker board solid phase~\cite{chenyang}.
Bosons with two-body onsite attractive interactions in optical lattices that are subjected to the onsite three- or four-body constraint
can form dimer and density wave phases as well ~\cite{vekua,daley1,daley2}.

\begin{figure}[t]
   \centering
\psfig{file=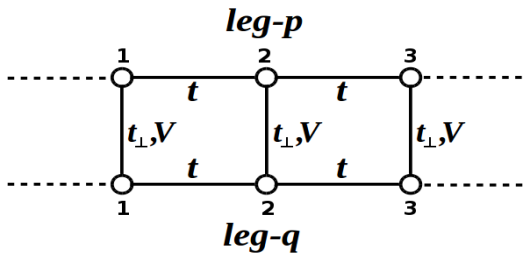,height=1.3in,width=2.7in,angle=0}
\caption{A small section of the ladder model is shown and the cluster contains six sites. 
Dashed lines represent bonds between decoupled sites,
circles represent lattice sites.}
\label{fig:fig1}
\end{figure}

In this paper, we focus on a system of ultracold bosonic atoms possessing long-range interactions along the rungs of a two leg ladder
as shown in Fig.~\ref{fig:fig1}.
This system can be realised by using dipolar bosons and polarizing them at the magic angle $\phi$ with reference to the plane of the ladder,
and perpendicular to the rungs of the ladder. The dipole-dipole interaction energy is proportional too $(1-3cos^2{\phi})$. Therefore if 
the angle $\phi$ is chosen such that the condition $cos^2{\phi}=1/3$ is satisfied,
the dipole-dipole interaction energy along the leg becomes zero but remains finite along the rungs.
Taking the onsite interactions to be attractive for such a system and imposing the three body constraint,
we study its ground state phase diagram. In addition, 
we also study the quantum phases of this model in the case of hard core bosons to validate our predictions.
We use the self consistent cluster mean-field theory (CMFT) to
determine the various order parameters to obtain the different quantum phase transitions in this model.

The remaining part of this paper is organized as
follows. In Sec. II we give details of the model and the method used in our calculation. Section III is devoted to
our results. Sec. IV contains concluding remarks. 

\section{Model and method}
The effective many-body lattice model which describes the problem mentioned above is given by
\begin{eqnarray}
%{\cal{H}}
H=-t\sum_{\alpha,i}(a_{\alpha,i}^{\dagger}a_{\alpha,i+1}+H.c.)-t_{\perp}\sum_{i} (a_{p,i}^{\dagger}a_{q,i}+H.c.)~~~~~~~ \nonumber \\
+\frac{U}{2}\sum_{\alpha,i}n_{\alpha,i}(n_{\alpha,i}-1)+V\sum_{i}{n_{p,i}n_{q,i}}-\sum_{i}\mu_{\alpha,i}{n_{\alpha,i}}~~~~~%\nonumber \\
\label{eq:ham}
\end{eqnarray}
where ${a_{\alpha,i}}^{\dagger} ({a_{\alpha,i}})$ is the bosonic creation (annihilation) operator at the site $i$ of
leg - ${ \alpha(=p,q)}$. $t$ is the hopping amplitude between the nearest neighbor sites along the legs of the ladder and
$t_\perp$ is the hopping along the rungs of the ladder.
${n}_{\alpha,i}={a}_{\alpha,i}^{\dagger}{a}_{\alpha,i}$ is the number operator at site ($\alpha,i$).
$U$ and  $V$ represent the on-site inter-atomic two-body and nearest-neighbor interactions respectively. The chemical potential
is represented by $\mu$. In this work we assume the three body constraint, \emph{i.e.} $(a^\dagger)^3|0\rangle=0$
and the value of $U$ is negative. We also assume that the ladder is arranged in such a way
that the nearest neighbor repulsion $V$ is only present along the rungs of the ladder.

The ground state properties of this model can be studied quite accurately by using powerful
numerical methods such as the density matrix
renormalization group method (DMRG)~\cite{white,scholwoek} or the quantum monte carlo (QMC) methods. However, in order to
qualitatively understand the quantum phase transitions exhibited in this model, we use the self consistent CMFT method. This method is capable
of capturing the relevant physics that arises due to quantum correlations, which was not always possible to achieve in the
conventional single site mean-field theory decoupling approximation~\cite{stoofmf,sheshadri,pai,jamshid,manpreet,arya}.
The CMFT can account for non-local interaction more accurately by retaining them in the exact form. 
For the regular SF-MI transition the estimation of the crtical point imporves when CMFT is used and approaches the value
obtained by the methods like QMC and DMRG ~\cite{macintosh}.
In this method a cluster of sites is treated exactly
and the mean-field approximation is for the coupling which connects the cluster with rest of the lattice.
The accuracy of the calculation depends on the size of the cluster considered.
As the cluster size increases, the number of decoupled bond reduces and the Hamiltonian approaches the exact form.
In one of its very first applications this method was used to study the phase diagram of a one-dimensional 
optical superlattice~\cite{vezzani}. In recent years it has been used to study various other models where the results 
are found to be in good agreement with those obtained using exact numerical methods~\cite{daniel,macintosh,danshitacmft,hassan}.

We consider six sites cluster as shown in Fig.\ref{fig:fig1}. Then the model given in Eq. (\ref{eq:ham}) can be written as
\begin{equation}\label{eq:ham2}
    H=\sum_k H^k_c
\end{equation}
where summation over $k$ is over all the clusters and $H^k_c$ is the Hamiltonian for $k^{th}$ cluster which can be written as
(dropping $k$ since all the clusters are identical)
\begin{equation}
H_{c}=H_{e}+H_{d},
\end{equation}
where $H_e$ is the exact Hamiltonian of the cluster and $H_d$ is the decoupled
hopping term to the nearest neighbor cluster. Using the mean-field decoupling approximation one can make the
following substitution since $i$ and $j$ are adjacent sites in the nearest neighbor clusters,
\begin{eqnarray}
{a_{i}^{\dagger}a_{j}}~\simeq~\langle{a_{i}^{\dagger}}\rangle{a_j}+
{a_i}^{\dagger}\langle{a_j}\rangle-\langle{a_i}^{\dagger}\rangle\langle{a_j}\rangle
\label{eq:two}
\end{eqnarray}
 in the hopping term in Eq.(1) and introduce the superfluid order parameter
\begin{eqnarray}
\phi_{i}\equiv\langle{a_{i}^{\dagger}}\rangle\equiv\langle{a_i}\rangle
\label{eq:three}
\end{eqnarray}
to obtain the following Hamiltonian for the decoupled part,
\begin{eqnarray}
{{H}}_{d}=-t\sum_{\alpha=\{p,q\}}~\sum_{\substack{i,i'=1,3\\i\neq{i'}}}\Big[\phi_{\alpha,i}({a_{\alpha,i'}}^{\dagger}+
a_{\alpha,i'})+\phi_{\alpha,i}\phi_{\alpha,i'}\Big]\nonumber \\
\label{eq:four}
\end{eqnarray}
The exact Hamiltonian, ${{H}}_{e}$ is given by
\begin{eqnarray}
{{H}}_{e}&=&-t\sum_{\substack{\alpha=\{p,q\}\\i=\{1,2\}}}(a_{\alpha,i}^{\dagger}a_{\alpha,i+1}+H.c.)\nonumber \\
&&-t_{\perp}\sum_{i=\{1,2,3\}}(a_{p,i}^{\dagger}a_{q,i}+H.c.) \nonumber \\
&&+\sum_{\substack{\alpha=\{p,q\} \\ i=\{1,2,3\}}}\Bigg[\frac{U}{2} n_{\alpha,i}(n_{\alpha,i}-1)+V{n_{p,i}n_{q,i}}-\mu_{\alpha,i}{n_{\alpha,i}}\Bigg]~~~~~
\label{eq:five}
\end{eqnarray}
We set the energy scale by choosing $t=1$, as a result, all the physical parameters considered are dimensionless.
We choose to work in the occupation number basis and construct the Hamiltonian matrix using the expression given
by $H_{c}$. This matrix is then diagonalized self-consistently to obtain the ground state of the system. The ground state
so obtained can then be used to calculate the necessary expectation values.
\section{Results and discussion}
In this section we report our findings and anlayze the results of our work.
Before presenting the main results, we validate our method; i.e. the CMFT
 by studying an already known phase diagram using other exact methods. It has been predicted that
in the case of hardcore bosons ($U=\infty$) on a ladder, the system exhibits a rung Mott insulator (RMI) phase due to
the competition between the intra-chain and inter-chain couplings, $t$ and $t_\perp$ respectively~\cite{carrasquilla,
simon}. When the value of $t_\perp$ is large compared to $t$, the atoms can only hop within the rungs of the
ladder, which results in the system being gapped. We have investigated this model using our CMFT approach and obtain
the phase diagram shown in Fig.\ref{fig:fig2}. This phase diagram is in a very good qualitative agreement with the
results presented in Ref.~\cite{simon}. 
We have done the calculations using $4-, 6-, 8-$ and $10-$sites clusters. The RMI$(1/2)$-SF critical transition point
found from each of these calculations are then scaled to the thermodynamic limit. 
This gives an estimate of $(t_{\perp}/t)_c\sim2.087$.
\bfig[!t]
  \centering
  \includegraphics*[width=0.415\textwidth,draft=false]{fig2.eps}
    \caption{(Color online) Phase diagram for hard-core bosons in absence of $V$ for different cluster sizes. The cluster sizes
    are indicated in the legend. The scaled critical point for RMI$(1/2)$-SF transition is represented by a black dot.}
    \label{fig:fig2}
\efig
\bfig[b]
  \centering
  \includegraphics*[width=0.47\textwidth,draft=false]{fig3.eps}
    \caption{(Color online) $\rho, \rho_s - \mu$ plot for (a)$U=-8.0$, (b)$U=-12.0$ and (c)$U=-15.0$ for $V=0.0$.
Solid (red) curves represent average density $\rho$ and circles (green curves) represent average
superfluid density $\rho_s$ in the cluster.}
    \label{fig:fig3}
\efig

We now turn to the results of our work.
Recent numerical and analytical works on the model given in Eq.(\ref{eq:ham}), in the absence of the nearest neighbor
interactions, have predicted the existence of a transition from the ASF phase to the DSF phase in one and two
dimensions~\cite{daley1,daley2,wessel,lee,chenyang}. Apart from this, two trivial gapped (insulating) phases,
MI(0) at $\rho=0$ and MI(2) at $\rho=2$, are also present.
It has been predicted for a two dimensional square lattice that (i) the MI to DSF transition is always continuous,
(ii) there exists a first order transition from the MI($0$) to ASF phase and (iii) there exists a tricritical point along the
ASF-MI($2$) transition boundary~\cite{wessel}.
In the presence of the nearest neighbor interaction $V$, the region of the first order phase boundary shrinks as $V$ is increased.
A dimer checker board solid (DCS) appears when $V\ne0$~\cite{chenyang}.
\bfig[t]
  \centering
  \includegraphics*[width=0.436\textwidth,draft=false]{fig4.eps}
    \caption{(Color online) Phase diagram for $U<0$ and $V=0.0$. Green (dashed) line and blue (dotted) line indicate
    first and second-order transitions, respectively. The first- to second-order change on the phase boundary is marked by a red circle.}
    \label{fig:fig4}
\efig

As stated earlier, the system we consider here is a two leg ladder and the nearest neighbor interaction is allowed only along
the rungs of the ladder.
We have considered three different cases: $V=0.0,~0.5$ and $1.0$ to demonstrate the salient features of phase diagrams for these three cases.
When $V=0.0$, the system exhibits a phase diagram qualitatively similar to that obtained for the
square lattice case as shown in Fig.\ref{fig:fig4}.
This phase diagram consists of four phases, the MI($0$), MI($2$), ASF and DSF phases.
To obtain this in our CMFT approach, we use the superfluid order parameter
$\phi$ and the density $\rho$ as the order parameters.
In the ASF phase, the atomic superfluid density $\rho_s=\phi ^2=\langle(a^\dagger)\rangle^2$
is finite and is zero for the MI phases. In other words the ASF phase is gapless and compressible
and the MI phases are gapped. In order to obtain the phase boundary between different phases
we plot the density $\rho$ (solid line (red curve)) and superfluid density $\rho_s$ (circles (green curve))
as a function of
the chemical potential $\mu$ for different values of $U$ in Fig.\ref{fig:fig3}
where it can be seen that the $\rho$ versus $\mu$ plot
(solid (red) curves) has two plateaus corresponding to the gapped MI(0) and MI(2) phases at $\rho=0$ and $\rho=2$
respectively. However, at intermediate densities the ASF and DSF phases appear. When $U$ is small, say $-8.0$, a region exists
where $\rho_s$ is finite, circles(green curve) in Fig.\ref{fig:fig3}(a) and the value of $\rho$ (solid(red) curves) increases
suddenly from zero to some finite value and then
increases continuously till it saturates at two. This region in the parameter space is
 the ASF phase since the superfluid
density $\rho_s$ remains finite. The sudden
jump in the values of $\rho$ and $\rho_s$ from zero to finite value suggests that the transition
from the MI($0$) to ASF phase is first order which will be discussed in more detail later. When $U$ becomes more negative and
the value of $\mu$ is small, the attraction between particles favors dimer formation which hop around to form
the DSF phase.
\bfig[!b]
  \centering
  \includegraphics*[width=0.47\textwidth,draft=false]{fig5.eps}
    \caption{(Color online) $U=-11.0, V=0.0$. (a)$\rho, ~\rho_s ~-~ \mu$ plot (b)single-particle tunneling and,
    (c)pair tunneling amplitude between sites.}
    \label{fig:fig5}
\efig
In the framework of our CMFT approach, it is difficult to predict the dimer phases directly from the calculation of
the DSF order parameter. However, if on increasing the value of the chemical potential, the
system density increases in steps of two atoms, then we conclude that it has entered the DSF phase.
This shows that the system consists of only dimers and they
behave as single entities. When the chemical potential is increased, for small values of $\mu$ as seen in the Fig.\ref{fig:fig3}(b)
for $U=-12.0$, the density of the system increases only if the chemical potential increases to accommodate two bosons or one dimer.
However, for higher values of the chemical potential, the density increases in a continuous manner with the chemical potential,
because the kinetic energy dominates in this situation and the system behaves like an atomic superfluid.
At this value of
$U$, a first order type transition from the ASF to the MI($2$) phase takes place, which can be seen from the sharp jump in the density at the
chemical potential close to the MI($2$) plateau. When $U$ is highly attractive, say $U=-15.0$,
all the particles form dimers and the system is fully in the DSF phase. The particle density then
 increases in steps of two particles till it reaches the MI($2$) state.
This behavior can be seen in the Fig.\ref{fig:fig3}(c). Since in this calculation
we have considered a cluster consisting of $6$ sites, we get jumps in the density when it reaches the
values $1/3,~2/3,~1,~ 4/3,~ 5/3$ and $2$ \emph{i.e.} when the total number of bosons in the cluster is equal to, respectively, $2,~4,~6,~8,~10$ and $12$.
 By locating the transition points from the $\rho~-~\mu$
curves we obtain various phases and the phase diagram as shown in Fig.\ref{fig:fig4}.
Although the DSF to ASF phase transition shows a first-order type behavior in the $\rho, \rho_s-\mu$ plots,
it is actually predicted to be of Ising type at unit filling in other models in earlier works\cite{daley1}. In our CMFT approach it is difficult to
predict the nature of this transition.
\bfig[!b]
  \centering
  \includegraphics*[width=0.47\textwidth,draft=false]{fig6.eps}
    \caption{(Color online) $U=-25.0, V=0.0$. (a)$\rho, ~\rho_s ~-~ \mu$ plot (b)single-particle tunneling and,
    (c)pair tunneling amplitude between sites.}
    \label{fig:fig6}
\efig
The small plateaus we obtain in $\rho-\mu$ curves in the DSF region are artifacts
of the finite size of the system we have used in our work. We expect these plateaus to become smaller and 
gradually disappear as the system size is increased. In fact, the disappearance of such plateaus has been shown in a comparative study of a system of hard-core bosons
using the exact-diagonalization and quantum Monte Carlo methods \cite{lauchli}.

Another signature of the dimer formation can be inferred by comparing the single-particle tunneling and the paired-tunneling
amplitudes. As an example we have plotted the above quantities for $U=-11.0$ and $U=-25.0$ in
Fig.\ref{fig:fig5} and Fig.\ref{fig:fig6} respectively.
The quantity $\langle{a_i}^\dagger{a_j}\rangle$ is the tunneling amplitude for a single boson
and $\langle{{a_i}^\dagger}^2{a_j}^2\rangle$ is the tunneling amplitude for a pair of bosons between the sites
$i$ and $j$. The pair of sites $i,j$ between which tunnelings are considered are given in the
legends of the respective plots. For $U=-11.0$, DSF exists only for a small region of $\mu$ values, around $\mu=-5.9$,
while the ASF phase dominates the rest of the region, as shown in Fig.\ref{fig:fig5}.
\bfig[!t]
  \centering
  \includegraphics*[width=0.42\textwidth,draft=false]{fig7.eps}
  \caption{(Color online) E($\phi$)-E($0$) versus order parameter ($\phi$) plot for $U=-9.0, V=0.0$.
   From top to bottom, $\mu=-6.0, -5.21, -5.114 ~(\mu_c), -5.05$ and $-4.95 $ for MI(0)-ASF transition across the left most
   boundary in Fig.\ref{fig:fig4}.}
\label{fig:fig7}
\efig
The paired-tunneling amplitude dominates over the single boson tunneling in the DSF phase and as expected both have finite values in the ASF phase.
When the DSF phase dominates, as in the case for $U=-25.0$, paired-tunneling amplitude
remains constant as $|i-j|$ increases while the single boson tunneling decreases to zero, as shown in Fig. \ref{fig:fig6}.
These features confirm our earlier conclusion that we do not have the ASF phase for higher values of $|U|$ as shown in Fig.\ref{fig:fig4}.
The order of the phase transition between MI(0) to ASF and ASF to MI(2) can be obtained by observing the ground state energy of the system
around the critical point on the common phase boundary.
We plot $E(\phi)-E(\phi=0)$ as a function of the superfluid order parameter $\phi$ for the $\mu$ values at and around the critical point.
One such plot for $U=-9.0, V=0.0$ is given in Fig.\ref{fig:fig7}.
This point lies on the phase boundary between
MI(0) and the ASF phases (Fig.\ref{fig:fig4}). When $\mu<-5.114$ the system is in the MI(0) phase
and we obtain a single
minimum, but as $\mu$ increases and approaches the critical point ($\mu_c=-5.114$), two more minima start appearing.
At the exact critical point all the three minima become degenerate. A single minimum indicates a unique solution which corresponds to
the MI phase and the three degenerate minima indicate the three of the possible solutions of the infinitely degenerate SF phase.
This is an indicator of a first-order transition.

\bfig[!t]
  \centering
  \includegraphics*[width=0.43\textwidth,draft=false]{fig8.eps}
  \caption{(Color online) E($\phi$)-E($0$) vs. order parameter ($\phi$) plot for $U=-5.0$.
From top to bottom, $\mu=1.21, 1.11, 1.01(\mu_c), 0.91$ and $0.81$  for ASF-MI(2) transition across the right most
boundary in Fig.\ref{fig:fig4}.}
\label{fig:fig8}
\efig
We pick one other point at $U=-5.0, V=0.0$ on the phase boundary between ASF and MI(2) phases and repeat the above
procedure to find out the order of transition. The corresponding plot is shown in Fig.\ref{fig:fig8}.
We can see that there are only two minima merging into a single minimum at this point. Therefore this is a second-order transition.
We repeat these calculations for several values of $U$ around which there are sudden jumps in $\rho, ~\rho_s-\mu$
plots. For $V=0.0$ we find that the nature of the SF-MI($2$) transition changes from first-order to second-order at $U\sim-9.4$. This point is
marked by a red circle in Fig.\ref{fig:fig4}, which is a tricritical point.

Now we discuss  our findings by considering the effect of the nearest neighbor interaction $V$. As
 mentioned before, we consider the nearest neighbor interaction only along the rungs of the
 ladder such that the system does not break any translational symmetry by forming
 a density wave order. In such a situation, the effect of a small value of $V$ is dramatic when
  $U$ is highly attractive. We investigate the system for two different values of $V$ equal to $0.5$ and $1.0$.
 In both the cases we study the phase diagram by increasing the magnitude of $U$ and making
 it more attractive. When $U$ is slightly negative the system exhibits the ASF phase for densities
 intermediate between $0$ and $2$. However, when the value of $U$ is sufficient to form dimers and at $\rho=1$, the small
\bfig[!t]
  \centering
  \includegraphics*[width=0.47\textwidth,draft=false]{fig9.eps}
    \caption{(Color online) $\rho, ~\rho_s-\mu$ plot for (a)$U=-8.0$, (b)$U=-10.5$ and (c)$U=-15.0$ for $V=0.5$ .
Solid (red) curves represent average density, solid-circle (green) curves represent average
superfluid density in the cluster.}
    \label{fig:fig9}
\efig 
\bfig[!b]
  \centering
  \includegraphics*[width=0.43\textwidth,draft=false]{fig10.eps}
    \caption{(Color online) Phase diagram for $U<0$ and $V=0.5$ showing different phases. 
Green (dashed) line and blue (dotted) line indicate
    first and second-order transitions, respectively.
The first- to second-order change on the phase boundary is marked by a red dot.}
    \label{fig:fig10}
\efig
\bfig[t]
  \centering
  \includegraphics*[width=0.43\textwidth,draft=false]{fig11.eps}
    \caption{(Color online) Phase diagram for $U<0$ and $V=1.0$ showing different phases. 
Green (dashed) line and blue (dotted) line indicate
    first and second-order transitions, respectively.
The first- to second-order change on the phase boundary is marked by a red dot. 
The change in phase boundaries with cluster size
is also indicated at $U=-9.0, -10.0$ and $-15.0$. For the 4-sites cluster, the DRI lobe does not extend beyond $U=-9.12$.}
    \label{fig:fig11}
\efig
 value of $V$ tries to prevent two dimers to sit on a single rung. However, it cannot
 restrict the dimer to hop within the sites of a rung which is governed by the kinetic term $t_\perp$.
 Hence the dimers are localized on the rungs of the ladder creating a singlet on each rung.
 This phase exhibits a finite single particle gap, and
 vanishing superfluid order parameter. This phase can be called as the dimer rung insulator (DRI). However,
 for the density range $0 < \rho < 1$ and $1 < \rho < 2 $, the system
 remains in the DSF phase. In order to obtain the phase diagram we analyze the plots of $\rho$ and $\rho_s$
 as a function of
 $\mu$ and is given for $V=0.5$ in Fig.\ref{fig:fig9}.
\bfig[b]
  \centering
  \includegraphics*[width=0.43\textwidth,draft=false]{fig12.eps}
  \caption{(Color online) $U=-25.0, V=1.0$. (a)$\rho, \rho_s-\mu$ plot (b)single particle tunneling amplitude between sites
  (c)pair tunneling amplitude between sites.}
    \label{fig:fig12}
\efig
 It is evident that when $U=-8.0$, the
 effect of $V$ is not visible as shown in Fig.\ref{fig:fig9}(a).
 However, Fig.\ref{fig:fig9}(b) shows that for $U=-10.5$ there appears a plateau at
 $\rho=1/3$ and $\rho=1$. The length of the $\rho=1/3$ plateau increases slightly but $\rho=1$
 plateau increases considerably as $|U|$ increases, as shown in Fig.\ref{fig:fig9}(c). At the
 $\rho=1$ plateau
region the value of $\rho_s$ is zero which reflects that the DRI phase is gapped. By picking the boundary
points from the $\rho$ versus $\mu$ curve we obtain the phase diagram as shown in Fig.\ref{fig:fig10}.
When the value of $V=1.0$, the DRI phase gets enlarged as shown in Fig.\ref{fig:fig11}.
Changes in the phase boundaries with a change in cluster size are also indicated in Fig.\ref{fig:fig11}.
A scaling of ASF-DRI critical point with 4-,6- and 8-sites clusters gives an estimate of $U_c\sim-7.92$ in the thermodynamic limit.
The order of the phase transitions, like before, are also obtained by simultaneously observing the sharp jump in the
corresponding $\rho-\mu$ plots and the quantity $E(\phi)-E(0)$. The position of the tricritical point
shifts to the higher values of $\mu/|U|$ as the value of $V$ increases. This phenomena was also predicted before in a similar model for a
square lattice~\cite{chenyang}.
\bfig[t]
  \centering
  \includegraphics*[width=0.47\textwidth,draft=false]{fig13.eps}
    \caption{(Color online) $\rho, \rho_s-\mu$ plot for HC bosons for (a)$V=0.0$ (b)$V=10.0$ and (c)$V=20.0$.}
    \label{fig:fig13}
\efig
The plateau at $\rho=1$ also appears in the DSF phase. In order to
distinguish between the DSF and DRI phases, we plot the single dimer correlation function (paired-tunneling amplitudes)
along the rung and the leg of the ladder.
When the system is in the DSF phase, this correlation function is finite both along the
rungs and the legs. However, in the DRI phase, it is large on the rungs compared to the legs. As the
value of $V$ increases, they tend to zero along the legs whereas they tend to one along the rungs as shown in
Fig.\ref{fig:fig12}. In Fig.\ref{fig:fig12}(a) we plot the $\rho, \rho_s - \mu$ plot for $U=-25.0$ and $V=1.0$
for comparison. In Fig.\ref{fig:fig12}(b) and (c) we plot the single particle and dimer correlations, respectively. 
At this density each rung has one dimer and is in a superposition of $|0,D\rangle$ and $|D,0\rangle$ states, where $D$ stands for a dimer.
It becomes energetically unfavorable for a dimer to hop from one rung to the another as the presence of $V$ will
tend to increase the energy. As a result dimers get confined to their respective rungs.
Therefore, we argue that the phase which appears at $\rho=1$ in the presence of $V$ is the DRI phase.
\bfig[b]
  \centering
  \includegraphics*[width=0.45\textwidth,draft=false]{fig14.eps}
    \caption{(Color online) Phase diagram for hard-core bosons in the presence of interchain nearest neighbour interaction $V$.}
    \label{fig:fig14}
\efig
The stability of the DRI phase in
the thermodynamic limit is difficult to predict using the CMFT, as it takes into account only 
a limited number of sites in a cluster. One needs to perform rigorous numerical calculations to understand this phase in 
more detail.

In order to further clarify the existence of the DRI phase, we study a model of hardcore bosons on a two leg
ladder with the nearest neighbor interactions acting along the rungs. This model at half filling is similar to
the model discussed above in the limit of large attractive $U$ at unit filling when all the atoms have formed dimers.
In the previous case, because of the
three body constraint the dimers behaved like hardcore bosons. Therefore, it is indeed possible to get the rung insulator (RI) phase in a
similar model of hardcore bosons.
The phase-diagram for hard-core bosons at $V=0.0$ is trivial and there are only SF and MI phases.
For low values of $\mu$ the system is in the $\rho=0.0$ MI phase. As $\mu$ is increased, the density of the system
increases continuously, it enters the SF phase and finally ends up in the $\rho=1.0$ MI phase.
However, by switching on the value of $V$, we obtain a plateau at $\rho=0.5$ which gets enlarged
as the value of $V$ increases, as shown in Fig.\ref{fig:fig13}.
The argument here is that, at $\rho=0.5$ and at finite $V$ the favorable ground state is when each rung of the
ladder has only one hardcore boson. In such a situation a singlet of hardcore bosons is formed along the rungs, which is
a rung Mott insulator phase as discussed before. We obtain the phase diagram for this model which
is shown in Fig.\ref{fig:fig14}. The presence of the RI phase can be further confirmed by comparing Fig.\ref{fig:fig15}(a) and
Fig.~\ref{fig:fig15}(b). We can see that in the former case when $V=0$, the tunneling amplitudes between all the sites
are almost the same for all values of $\mu$ but in the latter cases when $V\neq0$ they become different.
For $V\neq0$, at $\rho=0.5$ the tunneling amplitude of bosons within the same leg decreases and within a rung it increases.
\bfig[t]
  \centering
  \includegraphics*[width=0.45\textwidth,draft=false]{fig15.eps}
    \caption{(Color online) Single particle tunneling amplitudes for hardcore bosons at (a)$V=0.0$ and, (b)$V=20.0$.}
    \label{fig:fig15}
\efig

\section{Conclusions}
We have studied the phases and the phase transitions in an attractive Bose-Hubbard model on a two leg ladder in the presence
of the three body constraint. We obtain the ground state phase diagram of this model by using the self consistent
cluster mean-field theory. By calculating various physical parameters of interest, we find that there exists a
transition from the ASF to the DSF phases when the density of the system varies from zero to two. When the density is zero and two
we obtain two gapped phases such as MI($0$) and MI($2$). By
introducing nearest neighbor interactions between the particles sitting in the two sites of a rung, we obtain
the dimer rung insulator(DRI) phase at unit filling. The DRI phase is gapped in which the particle motion is
confined within the
rungs of the ladder. This phase appears in the middle of the DSF phase which gets enhanced as the value of the
nearest neighbor interaction increases.
We also find that the MI($0$)-ASF transition boundary is first order. However, the ASF-MI($2$)
phase boundary is continuous for small values of $|U|$ and becomes first order when $|U|$ is large through a
tricritical point. This point shifts towards the smaller values of $|U|$ as the value of $V$ increases.
We also complement our prediction of the DRI phase by studying a system of hardcore bosons
on a two leg ladder with nearest neighbor repulsions only along the rung.
To check the stability of the phases and scaling of the critical points we have done the calculations using 4, 6, 8 and 10
site clusters for the hardcore bosons. For the soft-core bosons with the three-body 
constraint we have done calculations upto 8-sites keeping $V$ fixed at $1.0$. In the case of hard-core bosons on a ladder we find that phase diagram improves with the increase in cluster size and
the RI-SF critical point approaches the value as obtained from DMRG and QMC calculations. In the case of softcore bosons with the
three-body constraint we find that overall the phase diagram remains the same qualitatively and there are only small changes 
in the phase boundaries with the change in cluster size.

\section{Acknowledgment}
We would like to acknowledge L. Santos, H. Weimer, M. Dalmonte, B. Laburthe-Tolra, S. Mukerjee,
A. Nunnenkamp, and D. Huerga for scientific discussions. The computational results reported in this work 
were performed on the high performance computing facilities of IIA, Bangalore. RVP thanks UGC (India) for support.

\end{document}